\documentclass[10pt,sigconf,letterpaper,prologue,table,natbib=false]{acmart}

\usepackage{tikz}
\usepackage{amsmath}
\usepackage{booktabs}
\usepackage{subcaption}
\usepackage{xcolor}
\usepackage{multirow}
\usepackage{balance}
\usepackage[plain]{fancyref}

\newcommand*{\fancyrefapplabelprefix}{app}
\Frefformat{plain}{\fancyrefapplabelprefix}{\MakeUppercase{\Freflstname}\fancyrefdefaultspacing#1#2}
\Frefformat{plain}{\fancyrefapplabelprefix}{Appendix~#1}

\usepackage[style=ACM-Reference-Format,%
            sorting=anyt,%
            maxbibnames=99,%
            mincrossrefs=1000,%
            backend=biber]{biblatex}

\renewcommand*{\bibfont}{\small}

\graphicspath{{figures/}}
\makeatletter
\def\input@path{{sections/}}
\makeatother

\definecolor{Gray}{gray}{0.9}

\title{Classifying Network Vendors at Internet Scale}

\author{Jordan Holland}
\affiliation{Princeton University}

\author{Ross Teixeira}
\affiliation{Princeton University}

\author{Paul Schmitt}
\affiliation{Princeton University}

\author{Kevin Borgolte}
\affiliation{Princeton University}

\author{Jennifer Rexford}
\affiliation{Princeton University}

\author{Nick Feamster}
\affiliation{University of Chicago}

\author{Jonathan Mayer}
\affiliation{Princeton University}

\begin{document}

\setcopyright{none}
\settopmatter{printacmref=false,authorsperrow=3} 
\renewcommand\footnotetextcopyrightpermission[1]{} 
\pagestyle{plain} 

\setlength{\abovedisplayskip}{4pt}
\setlength{\belowdisplayskip}{4pt}


\begin{abstract}
    In this paper, we develop a method to create a large, labeled dataset of visible network
    device vendors across the Internet by mapping
    network-visible IP addresses to device vendors. We use
    Internet-wide scanning, banner grabs of network-visible devices across the
    IPv4 address space, and clustering techniques to assign labels to more than 160,000 devices. We
    subsequently probe these devices and use features extracted from the
    responses to train a classifier that can accurately classify device
    vendors. Finally, we demonstrate how this method can be used to understand broader trends
    across the Internet by predicting device vendors in traceroutes
    from CAIDA's Archipelago measurement system and subsequently examining vendor
    distributions across these traceroutes.
\end{abstract}

\maketitle

\begin{sloppypar}
\section{Introduction}

Understanding the prevalence of different network device vendors can
lend insights into the robustness (or fragility) of the underlying
infrastructure. These devices run software that can contain
vulnerabilities~\cite{cisco2018routerbug,cisco2009routerbug,huaweirouterbug,cisconewrouterbug,yin2010towards,checkoway2016systematic},
and concerns have been raised about the possibilities of
``backdoors''~\cite{huaweibackdoor}.  Users and organizations may want better
ways to gain insights into the vendors of devices that are deployed in
different parts of the network; nations may also be interested in gathering
this intelligence on a broader scale.

Despite the benefits of fingerprinting network devices, doing so at
scale, remotely, is challenging. One challenge is gathering enough data to
develop models that can accurately classify devices by manufacturer.  Scanning
devices across the Internet is costly enough, but even given data from
Internet scans, the data lacks labels.

In this paper, we develop a method that can associate
network devices with vendors. The first step of the process involves
compiling a large, labeled dataset of more than 160,000 IP-visible network
devices across the Internet; for this part of the process, we use banner grabs
from SSH, Telnet, and SNMP to associate labels with corresponding
network-level devices. More specifically, we use a novel clustering approach to
extract vendor labels from banners. Second, we develop a probing technique that elicits
responses from these devices and can be used at scale, build a feature set based on these responses,
and train a classifier that predicts a vendor that corresponds to an IP
address with over 90\% accuracy. This second step is critical in allowing us to conduct a large-scale survey of
network devices: rather than only being able to classify devices that
have open SSH, Telnet, or SNMP ports, we can now classify {\em any} device
that responds to the measurement probes that we send.

The resulting measurements contain some sampling bias---in particular, the
resulting set only represents measurements of devices that are visible at the
network layer and that respond to the probes. Yet, assuming that any resulting
conclusions account for this bias, the set of measurements is
considerably larger than any existing dataset of its kind and allows
researchers to garner new insights from existing datasets.

Such a labeled dataset has considerable utility for the measurement community,
including the ability to augment existing datasets with these labels. We
demonstrate the insights that a model trained on a labeled dataset can yield by assigning labels to
network IP addresses in the CAIDA Archipelago traceroute dataset and exploring
the differences in the distribution of vendors on traceroutes from
the United States and Germany destined to different continents.

\section{\mbox{Generating Labeled Data at Scale}}
\label{sec:dataset}

\begin{figure*}[h]
\includegraphics[scale=.43]{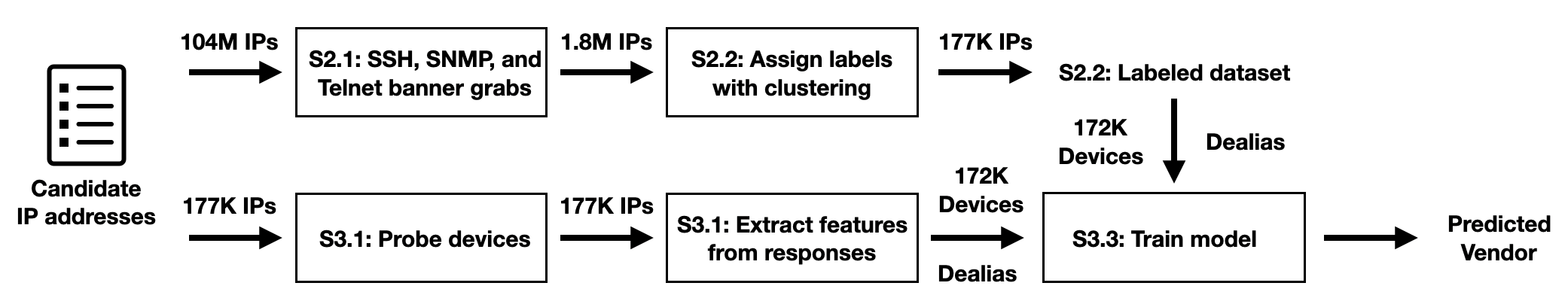}
\caption{Analysis pipeline.}
\label{fig:pipeline}
\end{figure*}

Figure \ref{fig:pipeline} shows our pipeline for classifying network devices at
scale. The first step in the pipeline involves creating a dataset that
associates IP
addresses with labels of different device vendors. As with many
supervised machine learning problems, acquiring or generating a large labeled
dataset can be a significant challenge. In this section, we present the method
we develop for generating
and curating such a dataset and describe the properties of the resulting
dataset.

More specifically, we combine clustering techniques with Telnet, Secure Shell
Protocol (SSH) and Simple Network
Management Protocol (SNMP) banners to aid in identifying the vendor of a
given device. Connecting to devices through these protocols can elicit a
banner, which can then be examined to determine the vendor of the device.
While we use the methods in this section to label network devices, the approach
can be generalized to other problem domains.

\subsection{Banner Grabs}

\textbf{Candidate IP Addresses.} As our goal is to classify network devices,
not end hosts, we rely on the CAIDA
Macroscopic
Internet Topology Data Kit (ITDK), which provides us with an IPv4 router
topology~\cite{caida-itdk}. The topology is produced from traceroutes from
December 24, 2019 to January 7, 2020 from 159 vantage points in 50 different
countries.\footnote{CAIDA's ITDK contains two topologies, one optimized for
accuracy of aliases and one optimized for coverage of aliases. We use the
topology optimized for accuracy.} The topology provides us with over 100
million IP addresses
associated with routers across the Internet, the ability to map each IP address
to its corresponding router, and a mapping from each network device to
the ASN controlling it.

We grab the banners for each of the over 100 million IP addresses in the IPv4
topology
using zgrab2~\cite{zgrab}. Zgrab2 supports SSH and Telnet banner grabs by
default, and we build a custom module for SNMP banner grabs.\footnote{SNMP
``banners'' are collected with a bulk query for “System” records.}  Table
\ref{tab:rr}
examines the magnitude and rate of
response of this process at different granularities. We see that although a
small percentage of devices respond to the requests, almost 2 million unique
devices respond to one of the banner grabs. Furthermore, these devices are
spread across over 18,000 ASes, accounting for over 30\% of the ASes we probed.

\subsection{Labeling }
Given the set of IP addresses and corresponding banners, we attempt to match
each IP address with a vendor.

\subsubsection{Vendor name matching}
We first try a simple, intuitive heuristic of matching an IP to a vendor label
by searching for vendor names
directly inside the banners. We collected a list of 40 network vendors from
Wikipedia~\cite{routerlist}. We
present the full list in Appendix \ref{app:regex}. Table \ref{tab:gt} shows
the results of this method. We
immediately see that Cisco dominates the results. Upon inspection, we find that
Cisco has a
branded SSH version which contributes to many of the labels.

This approach extracts an adequate amount of labels for some vendors, but can
miss
many phrases in the banners that may indicate the vendor of the router
without actually including the vendor's name. Furthermore, vendors
with more generic names will cause many conflicts or inaccurate labels using
this method. For
example, searching for ``extreme'' for Extreme Networks may include generic text
about ``extreme consequences for unauthorized connections.'' We encountered
168
conflicts using this approach where one banner matched multiple different vendor
names.

\subsubsection{Clustering.}
We now develop a more sophisticated approach to labeling IP addresses, based on
the
intuition that vendors configure devices with default banner formatting that is
often unique to a particular vendor. This default formatting may contain vendor
names, model numbers, warning messages, login prompts or other technical phrases
or sequences of characters that can be matched to a vendor through web searches.
By collecting these phrases, we are able to label significantly more devices for
many vendors and also discover new vendors.

We design an iterative clustering algorithm to search for potential matching
phrases or ``fingerprints.'' Clustering banners is a unique challenge that is
not easily
compatible with standard bag-of-words NLP models. Banner texts are overall
very similar, and features such as word ordering of arbitrary lengths and
whitespace can be very indicative. In addition, terms that
differ by one character (such as model numbers) may still be strongly related,
and the clustering should group them together. We therefore cluster banners
using their pairwise edit distances, which preserves banner structure.

We iteratively cluster random samples of unlabeled IP addresses, and use these clusters to generate
fingerprints. Take $N$ to be the dataset of banners. For
each
banner protocol, we repeatedly sample $M$ IP addresses from $N$ where $M \leq
1000$ (explained at the end of this section). For
each sample, we compute an $M$x$M$ symmetric edit distance matrix, where
\(M_{i,j} = M_{j,i}\) is the edit distance between banner $i$ and banner $j$. Specifically,
if \(len(i) \leq len(j)\), we compute $M_{i,j}$ by finding the Levenshtein edit
distances between $i$ and all $len(i)$-length substrings of $j$ and taking the minimum distance.

We then perform clustering on matrix $M$ for each banner protocol using HDBSCAN*~\cite{mcinnes2017hdbscan}.
We set $min\_cluster\_size$ and $min\_samples$ to 5, which causes HDBSCAN* to aggressively form small, tight clusters. HDBSCAN* also produces ``noise'' clusters for banners that do not strongly cluster, which we discard.

For each non-noise cluster, we identify potential fingerprints by computing the matching
sequences of characters between all combinations of banners and sorting them by frequency. We then link the most frequent
sequences to vendors by manually
searching on Google, Shodan, and Censys~\cite{durumeric2015a-serch-engine}. Possible links
include vendor names, model numbers, router documentation books, vendor support
forums and mailing lists, blog posts, and other banner protocols (e.g. FTP). For
example, \texttt{routeros ccr1} is from a MikroTik router family, and
\texttt{welcome to zxr} is from a ZTE router family.

If we encounter multiple similar fingerprints in a cluster, we manually create a
regex fingerprint to capture them. For example:
\[\verb+\(ttyp[\d\w]\)\r\u0000\r\n\r\u0000\r\nlogin+\] is used very frequently
by Juniper routers. If a fingerprint can be linked to multiple vendors or a
misleading resource, such as a textbook or enterprise customer, we ignore it. We
also create a ``blacklist'' for fingerprints linked to consumer routers or
endpoints, which we exclude from labeling.

After clusters of matrix $M$ have been examined, we apply our current set of fingerprints to all IP addresses and resolve conflicts that arise. Since we label the entire set
of IPs after each iteration of clustering, each iteration may significantly reduce the remaining sample
space of unlabeled IP addresses, so choosing sampling small matrices $M$ is sufficient.

\subsubsection{Conflict resolution}

Conflicting labels can arise when 1) a short
fingerprint matches too many banners as part of a longer sequence, or 2) devices include sequences from other vendors in their banners. If we can
confidently identify
the most likely vendor of a set of conflicts after additional web searches or close inspection, we set
that vendor's fingerprint to supersede the other. For example, we found SNMP banners
which
contained fingerprints for H3C or MikroTik as well as Huawei. On closer
inspection, we determined them to be H3C or MikroTik routers owned by Huawei, as
they also included H3C or MikroTik model numbers and OS versions.

If a fingerprint creates conflicts with more than one vendor, we remove the
fingerprint. For example: \[\verb+user access verification\r\n\r\nusername+\] is
a well-established Cisco Telnet banner fingerprint, and it appears in
\textit{Recog}, an open source fingerprint database~\cite{recog}. Clustering identified this fingerprint, and this fingerprint matched over 20,000 Cisco devices as well as devices with fingerprints from four other vendors.

\subsubsection{Results}
Ultimately, we are able to identify 30 different vendors and label over
175,000 IP addresses using our clustering approach, while our blacklist filters
out an additional 140,000 consumer or endpoint devices. Clustering
produced fewer than 20 conflicts throughout the entire process, compared to 168
conflicts from the vendor name matching approach, which provides confidence that
the process generates fingerprints with sufficiently low entropy. Table
\ref{tab:gt} shows the most frequently labeled
vendors.

We find many banners without vendor-specific phrases which cannot be assigned a
label. For example, 93\% of SSH banners only contain versions of Linux or
general SSH servers such as OpenSSH and Dropbear. Common reasons we were not able to associate banners with vendors were short authentication prompts (\texttt{login:} or \texttt{password}), client-side SSH
errors (\texttt{connection closed by remote host}) or empty banners.\footnote{Note that we treat all output from \texttt{zgrab} as banners. While that can include certain client-side prompts and errors, our approach to clustering filters those types of banners from our classification.} Lastly, banner
structures which appear very infrequently in our dataset are likely filtered out
by HDBSCAN* into the ``noise'' cluster.

Note that this approach is susceptible to introducing errors or failing to
detect conflicts, due to incomplete research, banner data, or fingerprints, as
well as devices which impersonate other vendor banners. For our
purposes, a small amount of noise in labels will have a negligible impact on
classifier performance. More broadly, despite these
potential pitfalls, our method provides the first fingerprint dataset targeting
network vendors learned from clustering and backed by web research, and we
show this is a promising approach for future label generation.

\begin{table}[t]
  \caption{Response rates to different banner grabs at multiple granularities.}\label{tab:rr}
  \resizebox{\columnwidth}{!}{%
    \begin{tabular}{@{}crrr@{}}
      \toprule
      Banner Type & \multicolumn{1}{c}{IP Addresses} & \multicolumn{1}{c}{Network Devices} & \multicolumn{1}{c}{ASes} \\ \midrule
      SSH & 1,399,147 (1.3\%) & 1,396,132 (1.3\%) & 13,791 (23.3\%) \\
      Telnet & 351,783 (0.3\%) & 349,519 (0.3\%) & 6,423 (10.9\%) \\
      SNMP & 181,732 (0.2\%) & 179,572 (0.2\%) & 7,030 (11.9\%) \\ \midrule
      {\bf Union} & {\bf 1,822,144 (1.7\%)} & {\bf 1,815,437 (1.7\%)} & {\bf 18,642 (31.4\%)} \\ \bottomrule
    \end{tabular}
}
\end{table}

\begin{table}[t]
  \caption{Labeled dataset. The first two columns represent labels generated from two different approaches. The last column represents the dataset resulting after removing unresponsive IP addresses and de-aliasing IP addresses to one device.}\label{tab:gt}
  \resizebox{\columnwidth}{!}{%
    \begin{tabular}{@{}lrrr@{}}
    \toprule
    \multicolumn{1}{c}{Manufacturer} & \begin{tabular}[c]{@{}c@{}}IP Address Labels:\\ Regex Match\end{tabular} &
        \begin{tabular}[c]{@{}c@{}}IP Address Labels:\\ Clustering\end{tabular} &
        \begin{tabular}[c]{@{}c@{}}Network Device Labels:\\ Responsive\end{tabular} \\ \midrule
    Cisco & 63,990 & 85,379 & 83,592 \\
    Mikrotik & 9,700 & 39,243 & 38,134 \\
    Huawei & 17,134 & 17,075 & 16,210 \\
    H3C & 10,231 & 10,620 & 10,183 \\
    NEC & 8 & 6,934 & 6,918 \\
    Lancom & 4,372 & 4,282 & 4,098 \\
    Juniper & 59 & 4,255 & 4,065 \\
    Adtran & 1,741 & 3,527 & 3,497 \\
    ZTE & 2,462 & 2,318 & 2,226 \\
    Ubiquoss & 8 & 1,887 & 1,869 \\
    Dell & 59 & 1,883 & 1,849 \\ \bottomrule
    \end{tabular}
  }
\end{table}

\begin{table}[t]
  \footnotesize
  \centering
  \caption{Mean classification performance using different sets of features across 5-fold cross validation.}\label{tab:ml-results}
  \begin{tabular}{@{}cccc@{}}
    \toprule
    Features                            & \begin{tabular}[c]{@{}c@{}}Balanced\\ Accuracy\end{tabular} & ROC AUC & F1   \\ \midrule
    Nmap                                & 77.3                                                        & 0.97   & 86.1 \\
    ICMP                                & 54.3                                                        & 0.91   & 73.7     \\
    \multicolumn{1}{l}{Nmap + Top ICMP} & 91.9                                                        & 0.99   & 93.7 \\ \bottomrule
  \end{tabular}
\end{table}

\section{Insights About Internet Paths}
\label{sec:insights}

To demonstrate the types of new insights that device classification can yield
for existing Internet measurement datasets, we collect a candidate set of network devices
that are likely active by parsing traceroutes from CAIDA's Archipelago (Ark)
infrastructure~\cite{caida-ark}, exploring  a snapshot of the Internet topology
by downloading and parsing Ark traceroutes from May 27,
2020. which yields network device IP addresses.
We collect a fingerprint for each IP address using Zmap, then
use the classifier trained in Section~\ref{sec:ml} to predict a class (i.e.,
vendor) for each
IP address. We determine the threshold for the classifier that provides
the best F1 score; if the classifier cannot predict a class with at least that
level of confidence, we assign an "unknown" label to the IP address.

\begin{table}[t]
  \footnotesize
  \caption{Distribution of the number of responses from each device in both our labeled dataset and the Internet-wide measurement.}\label{tab:zmap-rd}
  \centering
  \begin{tabular}{@{}lcrrrrrrrl@{}}
    \multicolumn{1}{c}{}        & \multicolumn{9}{c}{Number of Responses}
        \\ \cmidrule{2-10}
    \multicolumn{1}{c}{Dataset} & 0                        & \multicolumn{1}{c}{1} &
        \multicolumn{1}{c}{2} & \multicolumn{1}{c}{3} & \multicolumn{1}{c}{4} &
        \multicolumn{1}{c}{5} & \multicolumn{1}{c}{6} & \multicolumn{1}{c}{7} & 8
        \\ \midrule
    Labeled & \multicolumn{1}{r}{0.8}  & 5.7 & 4.2 & 7.7 & 3.6 & 7.3 & 23.6 & 46.7 & 0.6 \\
    Measurement & \multicolumn{1}{r}{23.9} & 2.5 & 9.9 & 15.1 & 5.3 & 4.8 & 13.1 & 22.0 & 3.4 \\ \bottomrule
  \end{tabular}
\end{table}

Table~\ref{tab:zmap-rd} shows that many more IP addresses
the CAIDA Ark dataset are unresponsive to probes than in the labeled dataset.
Table \ref{tab:zmap-rr} in Appendix \ref{app:census} shows the difference in
response rates for each specific probe in the labeled dataset and the CAIDA
measurement.

We categorize traceroutes based on the location of the source (i.e.,
country of CAIDA Ark node) and destination (i.e., continent of target IP
address, based on IP geolocation from \textit{geolite2})~\cite{geolite}. We
do not use IP geolocation on intermediate traceroute hops due to known
inaccuracy of geolocation on infrastructure IP
addresses~\cite{gharaibeh2017a-look}. For each
traceroute, we tally the unique vendors that are present and calculate the
probability of encountering each vendor over the dataset as: $
\frac{Traceroutes\ T\ with\ vendor\ V\ }{Total\ traceroutes\ T\ } $.
Table~\ref{tab:us_vs_de} shows the results for popular router vendors seen in
traceroutes from Arks located in the United States and Germany.

\begin{table}[t]
  \caption{Vendor prevalence in CAIDA Ark traceroute dataset, by source country and destination continent.}\label{tab:us_vs_de}
  \centering
  \resizebox{0.9\columnwidth}{!}{
    \begin{tabular}{llrrrr}
      \toprule
                          &               & \multicolumn{4}{c}{Vendor}             \\
      Source              & Destination   & Cisco   & Huawei  & Juniper & ZTE      \\
      \midrule
          \multirow{7}{*}{US} & Africa        & 80.9\% & 44.8\% & 81.3\% & 12.3\%  \\
                          & \cellcolor{Gray}Asia          & \cellcolor{Gray}84.5\% & \cellcolor{Gray}56.0\% & \cellcolor{Gray}76.8\% & \cellcolor{Gray}29.4\%  \\
                          & Europe         & 72.2\% & 37.0\% & 80.7\% & 25.5\%  \\
                          & \cellcolor{Gray}North America & \cellcolor{Gray}65.6\%
                          & \cellcolor{Gray}27.1\% & \cellcolor{Gray}73.1\% &
                          \cellcolor{Gray}19.0\%  \\
                          & Oceania       & 86.8\% & 35.4\% & 83.4\% & 19.4\%  \\
                          & \cellcolor{Gray}South America & \cellcolor{Gray}83.7\% & \cellcolor{Gray}43.2\% & \cellcolor{Gray}86.1\% & \cellcolor{Gray}25.8\%  \\
          & ALL           & 74.5\% & 39.4\% & 77.1\% & 23.7\%  \\
      \midrule
      \multirow{7}{*}{DE} & Africa        & 86.5\% & 58.0\% & 83.6\% & 37.2\%  \\
                          & \cellcolor{Gray}Asia          & \cellcolor{Gray}89.0\% & \cellcolor{Gray}69.9\% & \cellcolor{Gray}75.9\% & \cellcolor{Gray}61.5\%  \\
                          & Europe        & 78.4\% & 48.5\% & 72.1\% & 46.1\%  \\
                          & \cellcolor{Gray}North America & \cellcolor{Gray}88.1\% & \cellcolor{Gray}48.5\% & \cellcolor{Gray}68.6\% & \cellcolor{Gray}63.4\%  \\
                          & Oceania       & 91.1\% & 56.5\% & 81.0\% & 45.1\%  \\
                          & \cellcolor{Gray}South America & \cellcolor{Gray}91.3\% & \cellcolor{Gray}55.8\% & \cellcolor{Gray}85.8\% & \cellcolor{Gray}45.8\%  \\
                          & ALL           & 86.3\% & 55.7\% & 73.1\% & 56.7\%  \\
      \bottomrule
    \end{tabular}
  }
\end{table}

The source countries, United States and Germany, do show some differences in
vendor prevalence.
Traceroutes from sources in Germany are more likely to include Huawei
routers compared to traceroutes from sources in the United States.
Interestingly, prevalence of traceroutes from Germany to Europe
that Huawei is found on (48.52\%) is higher than that of traceroutes from the United States to
Europe (36.99\%). We also find that traceroutes from Germany
have a higher prevalence of ZTE devices, across all target continents. On the other hand,
Juniper is less prevalent on the traceroutes originating in Germany.

\section{Related Work}
\label{sec:related}

Vanubel et al. examined the feasibility of fingerprinting network equipment in
2013~\cite{vanaubel2013network-fingerprinting}. Specifically, they examine default
\textit{time to live} (TTL) headers in IP packets received by through active
probing. They create a signature for each router from the different initial
TTLs received from two separate ICMP requests. Our work considers TTL values as a
feature for the classification model, but vastly differs in the methods for
generating labels, classification, and types of probes sent to each
device. Further differentiating our work, we are the first to apply
machine learning techniques towards the discovery of network device vendors at
scale.

Feng et al. examine the automatic labeling of devices with a rule-based approach~\cite{feng2018acquisitional}.
At the vendor label, they look to label devices through the matching of 1,552 vendor names, specifically in the IoT space.
Our work differs in that we use clustering to discover more nuanced phrases that map to specific vendors, which we find produces a higher amount of labels with more confidence than regex matching.

Perhaps the most well known tool in active remote device fingerprinting is Nmap~\cite{nmap}.
Nmap is a network scanning tool that has developed into the most popular OS fingerprinting tool by using knowledge of the idiosyncrasies in the TCP/IP and the ICMP implementations in different types of devices.
Our work makes direct use of Nmap to actively probe devices.

Fingerprinting remote devices purely via ICMP messages was examined briefly by Arkin~\cite{arkin2002a-remote}.
Arkin developed X and XProbe to fingerprint remote operating systems solely through ICMP messages by examining the respective responses, like ICMP error message size and integrity.
Many of the techniques introduced by Arkin are now part of Nmap.

Kohno et al. introduced the area of remote \textit{physical} device fingerprinting
~\cite{kohno2005remote}. The work examined the feasibility of fingerprinting
remote devices by examining device clock skews. In this work, we do not
consider hardware fingerprinting methods, though we see it as a potential avenue
for future work in more granular device fingerprinting.

Raman et al. and Jones et al. leverage clustering techniques to develop
fingerprints~\cite{raman2020measuring,jones2014automated} for ground truth
labels. While our work uses clustering techniques to develop fingerprints in the
labeling process, we train models using the generated models on a different set
of features which allows us to fingerprint previously unlabeled devices at
scale.

\balance\label{lastpage}\section{Conclusion}
\label{sec:conclusion}

Understanding the manufacturers of network infrastructure on a given network is
valuable knowledge to multiple parties. In this paper, we explored the
feasibility of classifying network devices at scale. We have generated a
labeled dataset of over 160,000 devices using banner grabs, compiled a set of
probes that elicit unique responses from devices, trained a classifier to
distinguish between multiple vendors, and predicted network device vendors at
scale using this classifier. We specifically examine predicting IPv4 network
devices in this work, but believe the methodology could be generalized to IPv6
network devices or other device types, such as IoT devices.

\section*{Acknowledgments}
\label{sec:acknowledgments}

The authors thank Prateek Mittal and Liang Wang for valuable feedback on this work.
\end{sloppypar}

\newpage
\balance\printbibliography
\newpage
\clearpage
\appendix
\section{Appendix}
\label{sec:appendix}

\subsection{Regular Expression Match}
\label{app:regex}

Below are the list of manufacturers and simple regexes used to first match
banners to a manufacturer.

\begin{table}[b]
\begin{tabular}{@{}ll@{}}
\toprule
Manufacturer & Regexes            \\ \midrule
adtran       & adtran             \\
aerohive     & aerohive           \\
alaxala      & alaxala            \\
allied       & allied             \\
alcatel      & alcatel            \\
aruba        & aruba              \\
asus         & asus               \\
avaya        & avaya              \\
avm          & avm                \\
brocade      & brocade            \\
calix        & calix              \\
cisco        & cisco, "c i s c o" \\
dell         & dell               \\
draytek      & draytek            \\
d-link       & d-link             \\
enterasys    & enterasys          \\
ericsson     & ericsson           \\
extreme      & extreme            \\
fortinet     & fortinet           \\
h3c          & h3c                \\
hpe          & hpe, hewlett       \\
huawei       & huawei             \\
juniper      & juniper, junos     \\
lancom       & lancom             \\
linksys      & linksys            \\
meraki       & meraki             \\
mikrotik     & mikrotik           \\
netgear      & netgear            \\
nokia        & nokia              \\
openmesh     & open mesh          \\
ruckus       & ruckus             \\
sierra       & sierra             \\
technicolor  & technicolor        \\
tp-link      & tp-link, tplink    \\
trendnet     & trendnet           \\
ubiquiti     & ubiquiti           \\
xirrus       & xirrus             \\
yamaha       & yamaha             \\
zyxel        & zyxel              \\
zte          & zte,zhongxing      \\ \bottomrule
\end{tabular}
\end{table}

\subsection{Nmap Probes}
\label{app:nmap}

Here we further outline the 6 Nmap probes we used for fingerprinting routers.
First, we use 2 ICMP echo (ping) probes. The first ICMP echo probe sets the
don't fragment bit in the IP header to one, sets the IP type-of-service byte to
0, sets the ICMP code to 9 instead of 0, the ICMP sequence number to 295, and
sends 120 bytes of \texttt{0x00} for the paylaod. While Nmap randomly sets the
IP ID and ICMP request identifier, we use static values in our probes. The second ICMP echo probe
sets the IP type-of-service byte to 4, and the ICMP code to 0. It sends 150
bytes of \texttt{0x00} instead of 120. Finally, it increases the ICMP request
ID and sequence number by one from the first ICMP echo probe.

Nmap sends one UDP probe a closed port on the target machine. This UDP port
contains a payload of the character 'C' repeated 300 times. The IP ID value is
set to \texttt{0x1042}. Nmap sends this probe specifically to invoke an ICMP
port unreachable message from the target machine.

Finally, Nmap sends 3 TCP probes to a closed port on the target machine. The TCP
options for each probe are static, corresponding to a window scale of 10,
operation (NOP), a maximum segment size of 265, a timestamp value of
\texttt{0xFFFFFFFF}, and Selective ACKnowledgement (SACK) permitted. The
first TCP probe is a TCP SYN packet with a window field of 31337. The second TCP
packet is a TCP ACK packet with the don't fragment bit in the IP header set, and
a window field of 32768. The third TCP packet has the FIN, PSH, and URG flags
set and a window field of 65,535. Finally, the window scale option on this probe
is set to 15, rather than 10.

\subsection{Measurement Response Rate}
\label{app:census}

\begin{table}[h]
\centering
\caption{Response rates to each probe in both the labeled dataset and Internet-wide measurement.}
    \label{tab:zmap-rr}
  \resizebox{\columnwidth}{!}{%
    \begin{tabular}{@{}lrrrrrrrr@{}}
      & \multicolumn{8}{c}{Probe}\\
      \cmidrule{2-9}
      Dataset     & UDP  & \begin{tabular}[c]{@{}c@{}}ICMP\\ Echo 1\end{tabular} & \begin{tabular}[c]{@{}c@{}}ICMP\\ Echo 2\end{tabular} & \begin{tabular}[c]{@{}c@{}}ICMP\\ Timestamp\end{tabular} & \begin{tabular}[c]{@{}c@{}}ICMP\\ Address Mask\end{tabular} & TCP 1 & TCP 2 & TCP 3 \\ \midrule
      Labeled     & 61.6 & 88.3 & 96.7 & 70.7 & 13.2 & 80.9 & 77.8 & 72.5\\
      Measurement & 41.0 & 68.7 & 72.3 & 48.8 & 8.2 & 44.5 & 46.1 & 43.0\\
      \bottomrule
    \end{tabular}
  }
\end{table}

\subsection{Classification Features and Performance}
\label{app:class}

See table and figures on the following page.

\begin{table*}[h]
  \centering
  \caption{Features extracted from the responses to the sent probes, ranked by their importance to classification.}\label{tab:features}
  \resizebox{\textwidth}{!}{%
    \begin{tabular}{@{}llll@{}}
    \toprule
    Feature                                  & Description
        & Source & Rank \\ \midrule
        \rowcolor{Gray}
    IP initial time-to-live                  & Calculated original time-to-live of in response                                                                                                                     & Nmap   &     1       \\
    IP initial-time-to-live-guess            & If unable to calculate initial IP ttl, guess closest of \{32, 64, 128, 256\}                                                                                        & Nmap   &     2       \\
        \rowcolor{Gray}
    Responsiveness                           & Was the probe responded to?                                                                                                                                         & Nmap   &     3       \\
    Integrity of returned probe UDP checksum & Is the UDP header checksum value returned as it was sent?                                                                                                           & Nmap   &     4       \\
        \rowcolor{Gray}
    TCP RST data checksum                    & Is there error data in the TCP reset packet?                                                                                                                        & Nmap   &     5      \\
    TCP flags                                & Recorded TCP flags in response                                                                                                                                      & Nmap   &     6       \\
        \rowcolor{Gray}
    ICMP sequence number                     & ICMP sequence number received in response to ICMP packet                                                                                                            & ICMP   &     7       \\
    ICMP address mask                        & ICMP address mask received in response to ICMP address mask request                                                                                                 & ICMP   &     8       \\
        \rowcolor{Gray}
    ICMP identifier                          & ICMP identifier received in response to ICMP packet                                                                                                                 & ICMP   &     9       \\
    Returned probe ID value                  & Is the IP ID value in the response to the UDP probe the same as was sent?                                                                                           & Nmap   &     10       \\
        \rowcolor{Gray}
    ICMP code                                & ICMP code received in response to ICMP packet                                                                                                                       & ICMP   &     11       \\
    TCP window size                          & 16-bit window size in TCP header of response                                                                                                                        & Nmap   &     12       \\
        \rowcolor{Gray}
    TCP acknowledgement number               & Comparison of acknowledgement number in response to sequence number in probe                                                                                        & Nmap   &     13       \\
    ICMP echo response code                  & ICMP message codes received in response to Nmap ICMP echo packet                                                                                                    & Nmap   &     14       \\
        \rowcolor{Gray}
    TCP options                              & TCP options in response - order preserved                                                                                                                           & Nmap   &     15       \\
    IP don't fragment bit                    & Was the don't fragment bit set in the IP header?                                                                                                                    & Nmap   &     16       \\
        \rowcolor{Gray}
    TCP sequence number                      & Comparison of sequence number in response to acknowledgement number in probe                                                                                        & Nmap   &     17       \\
    IP total length                          & IP total length of response to UDP probe                                                                                                                            & Nmap   &     18        \\
        \rowcolor{Gray}
    Unused port unreachable field nonzero    & Are the last four bytes in the ICMP port unreachable message set?                                                                                                   & Nmap   &     19       \\
    Integrity of returned UDP data           & Is the UDP data returned exactly as it was sent?                                                                                                                    & Nmap   &     20       \\
        \rowcolor{Gray}
    ICMP type                                & ICMP message type received in response to ICMP packet                                                                                                               & ICMP   &     21       \\
    TCP quirks                               & \begin{tabular}[c]{@{}l@{}}Is the reserved field in the TCP header of response nonzero? \\ Is the urgent pointer field non-zero when URG flag not set?\end{tabular} & Nmap   &     22       \\ \bottomrule
    \end{tabular}
  }
\end{table*}

\begin{figure*}[h]
    \centering
    \begin{subfigure}[b]{0.95\columnwidth}
      \centering
        \includegraphics[width=\linewidth]{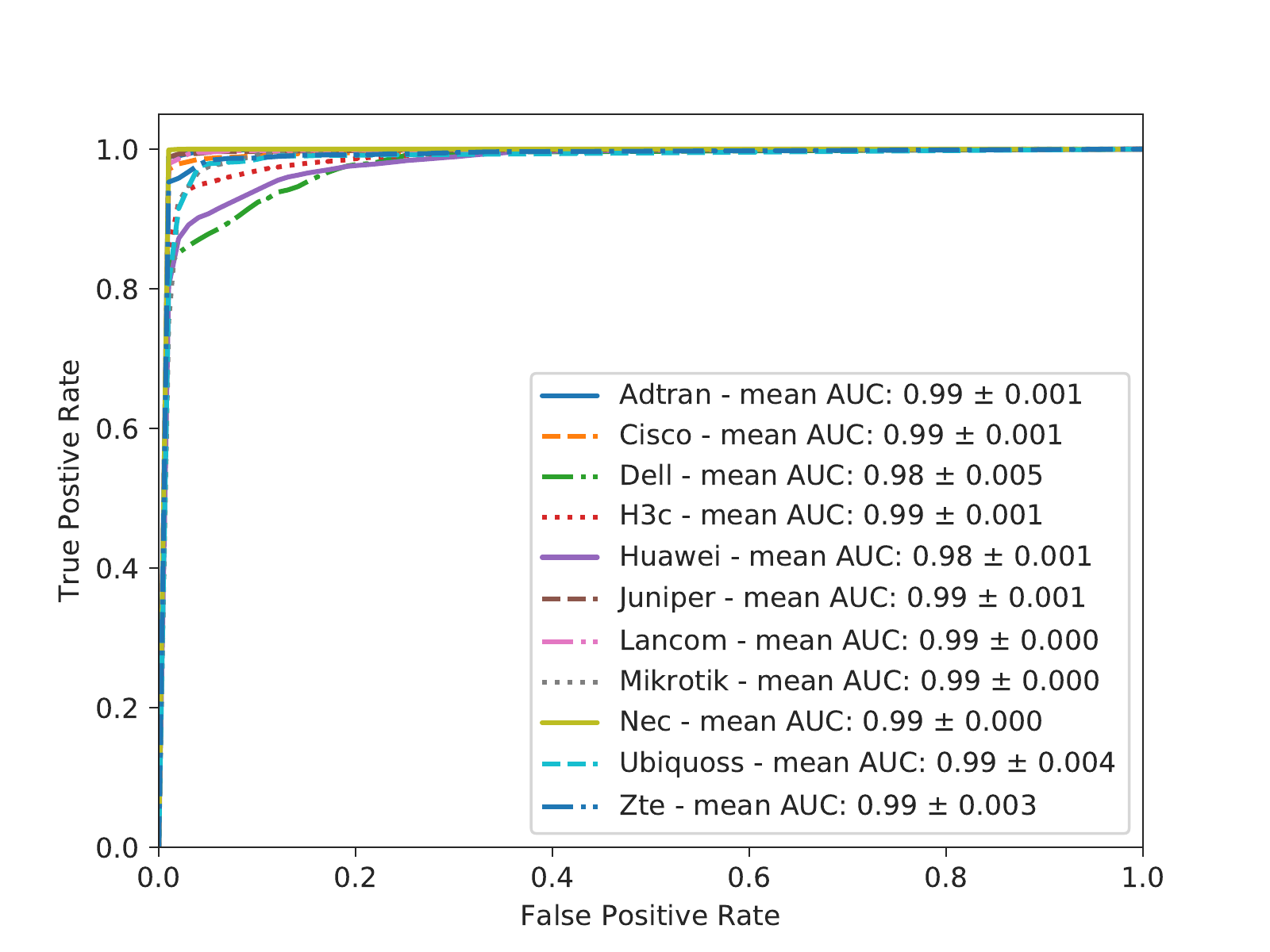}
      \label{fig:nmap_router_roc}
        \caption{ROC curve for our final trained model.}
    \end{subfigure}
    \begin{subfigure}[b]{0.95\columnwidth}
      \centering
      \includegraphics[width=\linewidth]{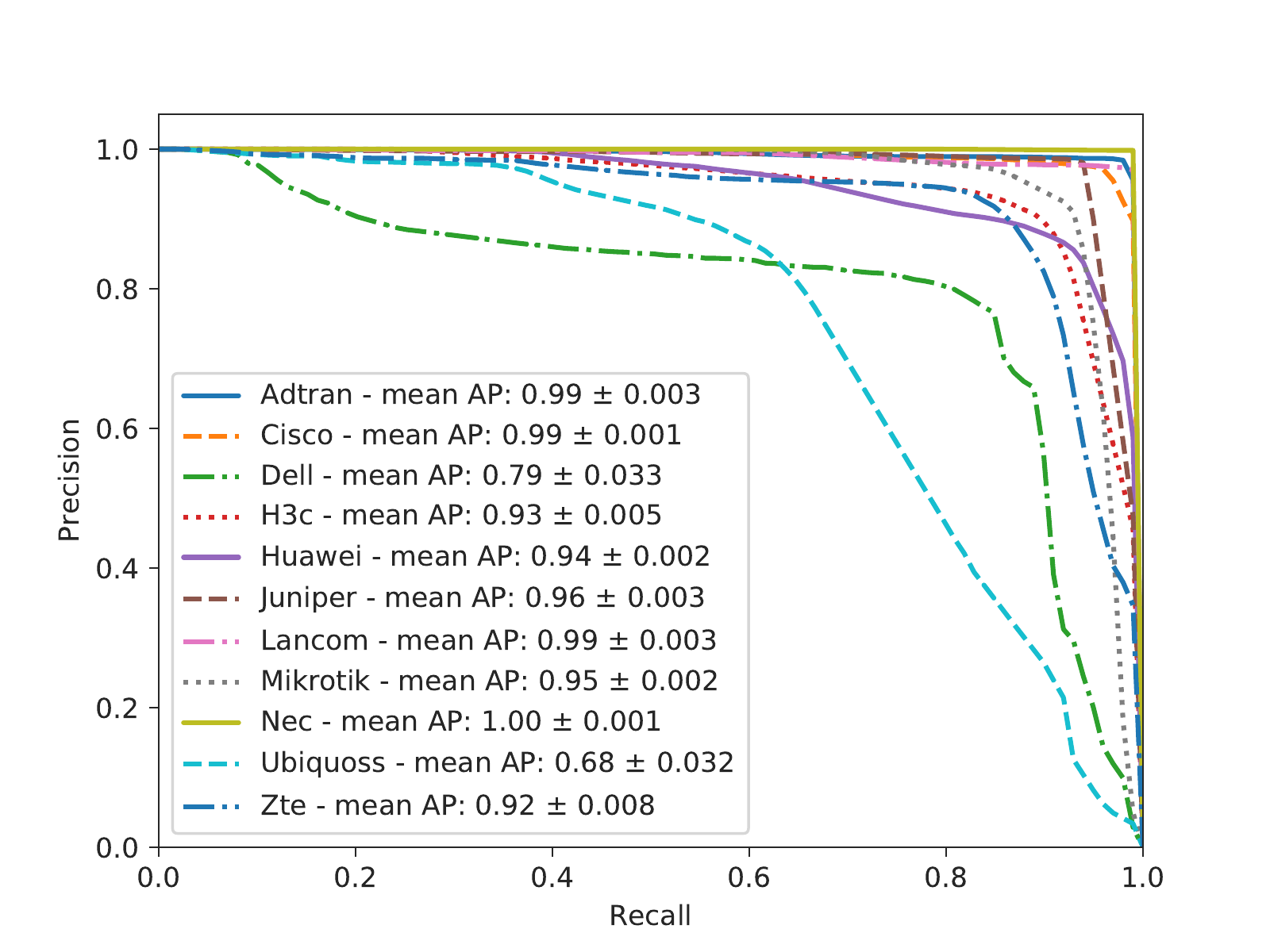}
      \label{fig:nmap_router_pr}
        \caption{PR curve for our final trained model.}
    \end{subfigure}
    \caption{Visualizing classification performance for different vendors.}
    \label{fig:rocpr}
\end{figure*}

\end{document}